\documentclass[twocolumn,showpacs,aps,prl,superscriptaddress,preprintnumbers,amsmath,amssymb,letter,floatfix]{revtex4}

\usepackage{colordvi}
\usepackage{graphicx}
\usepackage{dcolumn}
\usepackage{bm}
\usepackage{multirow}

\input babarsym
\def\Rdcs       {\ensuremath{R_{\rm D}}\xspace} 
\def\RdcsPM     {\ensuremath{R_{\rm D}^{\pm}}\xspace}

\def\xPrimeSq   {\ensuremath{{x^{\prime}}^2}\xspace}
\def\yPrimeSq   {\ensuremath{{y^{\prime}}^2}\xspace}
\def\xPrime     {\ensuremath{x^{\prime}}\xspace}

\def\yPrime     {\ensuremath{y^{\prime}}\xspace}
\def\yPrimeP    {\ensuremath{y^{\prime+}}\xspace}
\def\yPrimeM    {\ensuremath{y^{\prime-}}\xspace}
\def\xPrimePmSq {\ensuremath{{x^{\prime}}^{2\pm}}\xspace}

\def\yPrimePm   {\ensuremath{y^{\prime\pm}}\xspace}

\def\xPrimePSq  {\ensuremath{{x^{\prime}}^{2+}}\xspace}
\def\xPrimeMSq  {\ensuremath{{x^{\prime}}^{2-}}\xspace}
\def\xPSQyP     {\ensuremath{(\xPrimeSq,\,\yPrime)}\xspace}

\def\AD         {\ensuremath{A_{\rm D}}\xspace}

\def\pisoft     {\ensuremath{\pi_{\rm s}}\xspace}

\def\Dztokpi    {\ensuremath{\Dz \to K^{-}\pi^{+}}\xspace}

\def\Dzbtokpi   {\ensuremath{\Dzb \to K^{+}\pi^{-}}\xspace}
\def\DztokpiWS  {\ensuremath{\Dz \to K^{+}\pi^{-}}\xspace}

\def\Kmpip      {\ensuremath{K^{-}\pi^{+}}\xspace}
\def\Kmp        {\ensuremath{K^{\mp}}\xspace}

\def\pimp       {\ensuremath{\pi^{\mp}}\xspace}
\def\pipm       {\ensuremath{\pi^{\pm}}\xspace}

\def\mDstp      {\ensuremath{m_{\Dstarp}}\xspace}

\newcommand{\kevcc}{\ensuremath{{\mathrm{\,Ke\kern -0.1em V\!/}c^2}}\xspace}
\def\mKpi       {\ensuremath{m_{K\pi}}\xspace}
\def\dm         {\ensuremath{\Delta m}\xspace}
\def\t          {\ensuremath{t}}
\def\terr       {\ensuremath{\sigma_{\t}}}
\def\chisq       {\ensuremath{\chi^2}}
\def\Pchisq      {\ensuremath{P(\chi^2)}}

\def\lumitotnum  {\ensuremath{384}}  
\def\lumitot     {\ensuremath{\lumitotnum\invfb}}

\def\Dstp        {\ensuremath{D^{*+}}}

\def\terr   {\ensuremath{\delta t}\xspace}

\def\like   {\ensuremath{{\cal L}}}
\def\signif {\ensuremath{{s}}}

\def\Tws  {\ensuremath{T_{\rm WS}(t)}}

\def\mdmnospace            {\ensuremath{\{\mKpi,\,\dm\}}}
\def\mdm            {\mdmnospace\xspace}

\def\Pbar    {\ensuremath{\kern 0.2em\overline{\kern -0.2em P}{}}\xspace}

\def\G       {\ensuremath{\Gamma}\xspace}

\def\Dm      {\ensuremath{\Delta m}\xspace}
\def\DM      {\ensuremath{\Delta M}\xspace}
\def\DG      {\ensuremath{\Delta\Gamma}\xspace}

\def\Di      {\ensuremath{D_1}\xspace}
\def\Dii     {\ensuremath{D_2}\xspace}
\def\Mi      {\ensuremath{M_1}\xspace}
\def\Mii     {\ensuremath{M_2}\xspace}
\def\Gi      {\ensuremath{\Gamma_1}\xspace}
\def\Gii     {\ensuremath{\Gamma_2}\xspace}

\catcode`\@=11\relax
\newskip\dkwidth
\def\dk{%
   \dkwidth=2em plus 0.5 em minus 0.25 em\relax
   {\m@th\mathord{%
   \hbox{%
      \kern 0.3em
      \raise 0.6ex%
      \hbox{%
         \vrule width 0.25pt height 0.5\dkwidth depth0pt}%
      \kern-1.2pt%
      \hbox to 1.1\dkwidth{%
         \rightarrowfill}%
      \kern0.4em}}%
   }%
}%
\def\rightarrowfill{$\m@th\mathord-\mkern-10mu%
  \cleaders\hbox{$\mkern-2mu\mathord-\mkern-2mu$}\hfill
  \mkern-6mu\mathord\rightarrow$}
\catcode`\@=12\relax
\newcommand{\BABARPubYear}    {07}
\newcommand{\BABARPubNumber}  {019}

\newcommand{\SLACPubNumber} {12385}

\iffalse 

\else

\fi

\long\def\inst#1{\par\nobreak\kern 4pt\nobreak
    {\it #1}\par\vskip 10pt plus 3pt minus 3pt}

\begin{document}

\preprint{\babar-PUB-\BABARPubYear/\BABARPubNumber} 
\preprint{SLAC-PUB-\SLACPubNumber}

\title{Evidence for {\boldmath$\Dz$-$\Dzb$} Mixing}

%
\author{B.~Aubert}
\author{M.~Bona}
\author{D.~Boutigny}
\author{Y.~Karyotakis}
\author{J.~P.~Lees}
\author{V.~Poireau}
\author{X.~Prudent}
\author{V.~Tisserand}
\author{A.~Zghiche}
\affiliation{Laboratoire de Physique des Particules, IN2P3/CNRS et Universit\'e de Savoie, F-74941 Annecy-Le-Vieux, France }
\author{J.~Garra~Tico}
\author{E.~Grauges}
\affiliation{Universitat de Barcelona, Facultat de Fisica, Departament ECM, E-08028 Barcelona, Spain }
% ...
\collaboration{The \babar\ Collaboration}
\noaffiliation

\date{\today}

\begin{abstract}
We present evidence for $\Dz$-$\Dzb$ mixing 
in \DztokpiWS decays from \lumitot\ of \epem colliding-beam data 
recorded near $\sqrt{s}=10.6\gev$ with
the \babar\ detector at the PEP-II storage rings at SLAC.
We find the mixing parameters 
$\xPrimeSq = [ -0.22 \pm 0.30 \hbox{ (stat.)}\pm 0.21 \hbox{ (syst.)}]\times 10^{-3}$ 
and 
$\yPrime = [9.7 \pm 4.4 \hbox{ (stat.)}\pm 3.1 \hbox{ (syst.)}] \times 10^{-3}$,
and a correlation between them of $-0.94$. This result is
inconsistent with the no-mixing hypothesis with a 
significance of 3.9 standard deviations.
We measure \Rdcs, the ratio of
doubly Cabibbo-suppressed to Cabibbo-favored decay rates,
to be $[0.303\pm0.016\hbox{ (stat.)}\pm 
0.010\hbox{ (syst.)}]\%$.
We find no evidence for \CP violation.

\end{abstract}

\maketitle


\par
We present evidence for charm-meson (\Dz-\Dzb)
mixing.  This work complements results 
in the neutral $K$~\cite{Lande:1956pf,Fry:1956pg},
$B$~\cite{Albajar:1986it,Albrecht:1987dr},
and $\Bs$~\cite{Abazov:2006dm,Abulencia:2006ze} systems.
Although precise predictions are difficult, \Dz-\Dzb mixing 
in the Standard Model (SM) is expected at the 
1\% level or less~\cite{Petrov:2006nc,Falk:2004wg,Burdman:2003rs,Falk:2001hx,Donoghue:1985hh,Wolfenstein:1985ft}.
Our result is consistent 
with this expectation
and previous experimental 
limits~\cite{Aitala:1996fg,Zhang:2006dp,Godang:1999yd,Link:2004vk,Aubert:2003ae,Abulencia:2006sz}.
By observing the wrong-sign decay 
\DztokpiWS~\cite{footnote:CC},
we determine \Rdcs, the ratio of doubly Cabibbo-suppressed 
to Cabibbo-favored decay rates, and the mixing parameters $\xPrimeSq$
and $\yPrime$.

\par
The charm sector is the only place where the contributions to
\CP violation of down-type 
quarks in the mixing diagram can be explored.
We compare results for \Dz and \Dzb decays but find no evidence for 
\CP violation.
The SM predicts
effects well below the sensitivity of this
experiment.

\par
We study 
the right-sign (RS), Cabibbo-favored (CF)
decay $\Dztokpi$ and the
wrong-sign (WS) decay $\DztokpiWS$.  The latter can be produced via
the doubly Cabibbo-suppressed (DCS) decay $\DztokpiWS$ 
or via mixing followed by a CF
decay $\Dz \rightarrow \Dzbtokpi$.
The DCS decay has a small rate \Rdcs 
of order $\tan^4\theta_C \approx 0.3\%$ relative to CF decay.
We distinguish \Dz and \Dzb by their production in the 
decay $\Dstp\to\pisoft^+\Dz$ where the 
$\pisoft^+$ is referred to as the ``slow pion''. 
In RS decays the $\pisoft^+$ and kaon have opposite charges,
while in WS decays the charges are the same.
The time dependence of the WS decay rate
is used to separate the contributions of DCS decays from \Dz-\Dzb mixing.

\par
The \Dz and \Dzb mesons are
produced as flavor eigenstates,
but evolve and decay as mixtures of the eigenstates \Di and \Dii 
of the Hamiltonian, with masses and widths
\Mi, \Gi and \Mii, \Gii, respectively. 
Mixing is characterized by the mass and lifetime
differences $\DM = \Mi - \Mii$ and $\DG = \Gi - \Gii$.  Defining
the parameters $x = \DM/\Gamma$ and $y = \DG/2\G$, where 
$\G = (\Gi + \Gii)/2$, we approximate the time dependence of the WS decay
of a meson produced as a \Dz at time~$t = 0$ 
in the limit of small mixing ($|x|$, $|y| \ll 1$) and \CP 
conservation as 
\begin{equation}
    \frac{\Tws}{e^{-\Gamma t}} \propto
      \Rdcs + 
      \sqrt{\Rdcs}\yPrime\; \Gamma t + 
      \frac{\xPrimeSq + \yPrimeSq}{4} (\Gamma t)^2\,,
\label{eq:Tws}
\end{equation}
where $\xPrime = x\cos\delta_{K\pi} + y \sin\delta_{K\pi}$,
$\yPrime = -x\sin\delta_{K\pi} + y \cos\delta_{K\pi}$, and 
$\delta_{K\pi}$ is the strong phase between the DCS and CF amplitudes.
\par
We study both \CP-conserving and \CP-violating cases.
For the \CP-conserving case, we fit for the parameters \Rdcs, \xPrimeSq, and \yPrime. To search for \CP violation,
we apply Eq.~(\ref{eq:Tws}) 
to \Dz and \Dzb samples separately, fitting for the parameters
\{$\RdcsPM$, $\xPrimePmSq$, $\yPrimePm$\} for \Dz($+$) decays and 
\Dzb($-$) decays.

\par
We use  
$\lumitot$ of \epem\ colliding-beam 
data recorded near $\sqrt{s} = 10.6\gev$
with the \babar\ detector~\cite{Aubert:2001tu} at the
PEP-II asymmetric-energy storage rings.
We select \Dz candidates by pairing  
oppositely-charged tracks with a $\Kmp\pipm$ invariant
mass \mKpi between $1.81$ and $1.92\gevcc$,
requiring each track to have at least 12~coordinates in 
the drift chamber (DCH). Each pair is identified as $\Kmp\pipm$
using a likelihood-based particle identification 
algorithm.  The identification
efficiency for kaons (pions) is about $85\%$ $(95\%)$; the 
misidentification rate of kaons (pions) as pions (kaons) is about
$2\%$ $(6\%)$. 
\par
To obtain the proper decay time \t\ and its error~\terr\ for each
\Dz candidate, we refit the \Kmp and \pipm tracks, constraining
them to originate from a 
common vertex. We also require the \Dz and $\pisoft^+$ 
to originate from a common vertex, 
constrained by the position and size of the \epem interaction region. 
We require the $\pisoft^+$ to have a momentum  in the laboratory frame
greater than $0.1\gevc$ and in the \epem center-of-mass (CM) frame
below $0.45\gevc$.
We require the \chisq\ probability of the vertex-constrained combined fit 
$\Pchisq$ to be at least $0.1\%$, and the $\mDstp-\mKpi$ mass difference \dm to satisfy
$0.14<\dm<0.16\gevcc$.
\par
To remove \Dz candidates from $B$-meson decays and to reduce
combinatorial backgrounds, we require
each \Dz to have a momentum in the CM frame greater than
$2.5\gevc$. We require $-2 < \t<4\ps$
and $\terr<0.5\ps$
(the most probable value of \terr for signal events is $0.16\ps$).
For \Dstp\ candidates sharing one or more tracks with other \Dstp\ candidates,
we retain only the candidate with the highest \Pchisq.
After applying all criteria, we keep approximately 
1,229,000~RS and 64,000~WS
\Dz and \Dzb candidates.
To avoid potential bias, we finalized our data selection criteria 
and the procedures for fitting and extracting the statistical limits
without examining the mixing results.

The mixing parameters are determined in an unbinned, extended
maximum-likelihood fit to the RS and WS data samples over the four
observables \mKpi, \dm, $t$, and \terr. The fit is performed in several
stages. First, RS and WS signal and
background shape parameters are determined from a fit to \mKpi and
\dm, and are not varied in subsequent fits.  Next, the
\Dz proper-time resolution function and lifetime  are
determined in a fit to the RS data using \mKpi and \dm to separate the
signal and background components. We fit to the WS data sample using
three different models. The first model assumes both \CP conservation
and the absence of mixing, and only measures \Rdcs. The second model
allows for mixing, but assumes no
\CP violation, and the third model allows for both mixing and \CP violation.

The RS and WS \mdm\ distributions are described by four components:
signal, random $\pisoft^+$, misreconstructed \Dz and combinatorial
background. Signal has a characteristic peak in both \mKpi and \dm.
The random $\pisoft^+$ component models reconstructed \Dz decays
combined with a random slow pion and has the same shape
in \mKpi as signal events, but does not peak in \dm.  Misreconstructed
\Dz events have one or more of the \Dz decay products either not
reconstructed or reconstructed with the wrong particle
hypothesis. They peak in \dm, but not in \mKpi. For RS events, most of
these are semileptonic decays $\Dz\to\Km\ell^+\nu$ with the charged
lepton misidentified as a pion.  For WS events, the main contributor
is RS $\Dz\to\Kmpip$ decays where the
\Km and the \pip are misidentified as \pim and \Kp, respectively. 
Combinatorial background events are those not described by the
above components; they do not exhibit any peaking structure in
\mKpi or \dm.
\par
The functional forms of the probability density functions (PDFs) for
the signal and background components are chosen based on studies of
Monte Carlo (MC) samples. However, all parameters are determined from
two-dimensional likelihood fits to data over the full
$1.81<\mKpi<1.92\gevcc$ and $0.14<\dm<0.16\gevcc$ region.

\begin{figure}[phtb]
  \centering
  \centerline{%
    \includegraphics[width=0.5\linewidth, clip=]{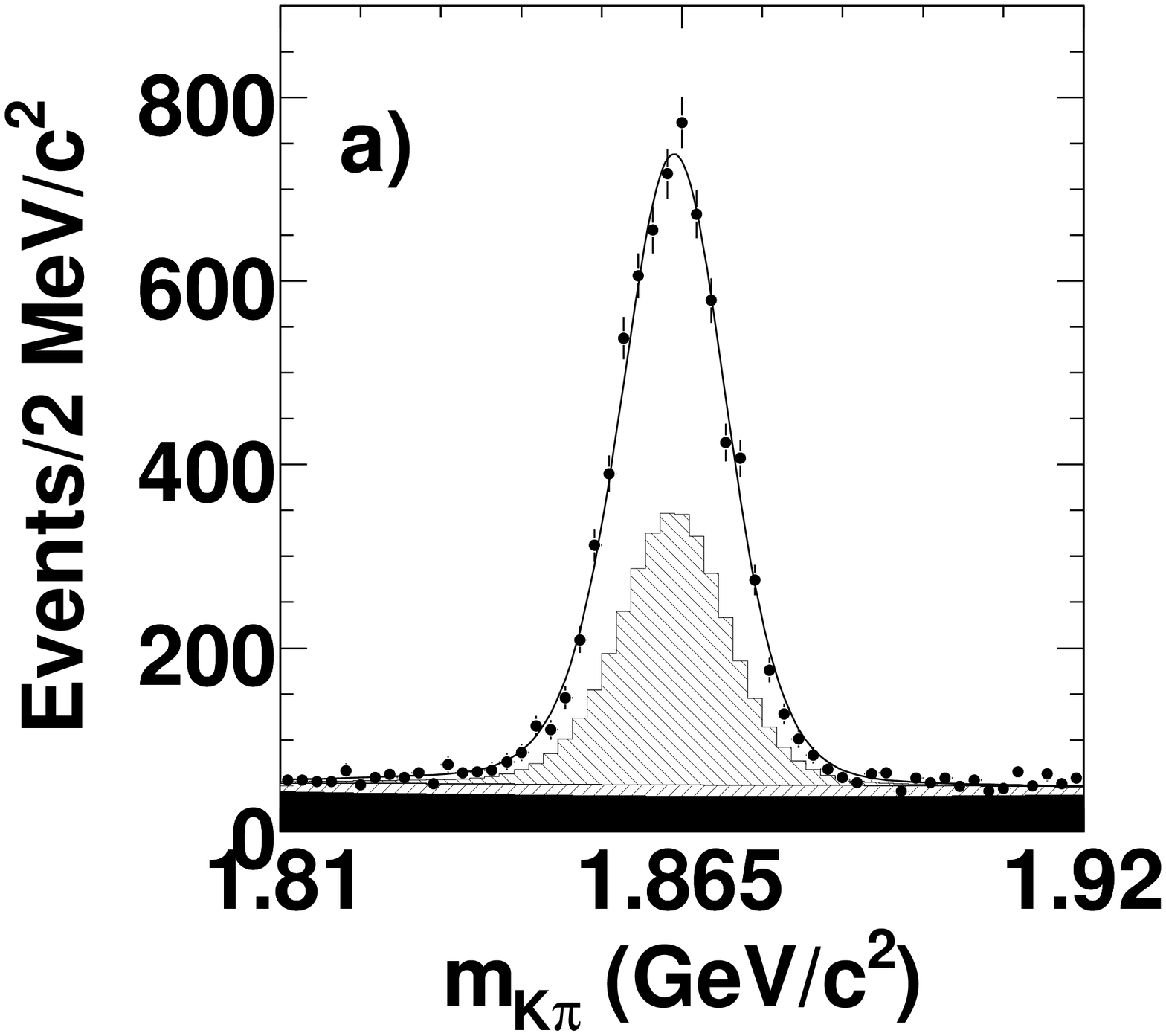}
    \includegraphics[width=0.5\linewidth, clip=]{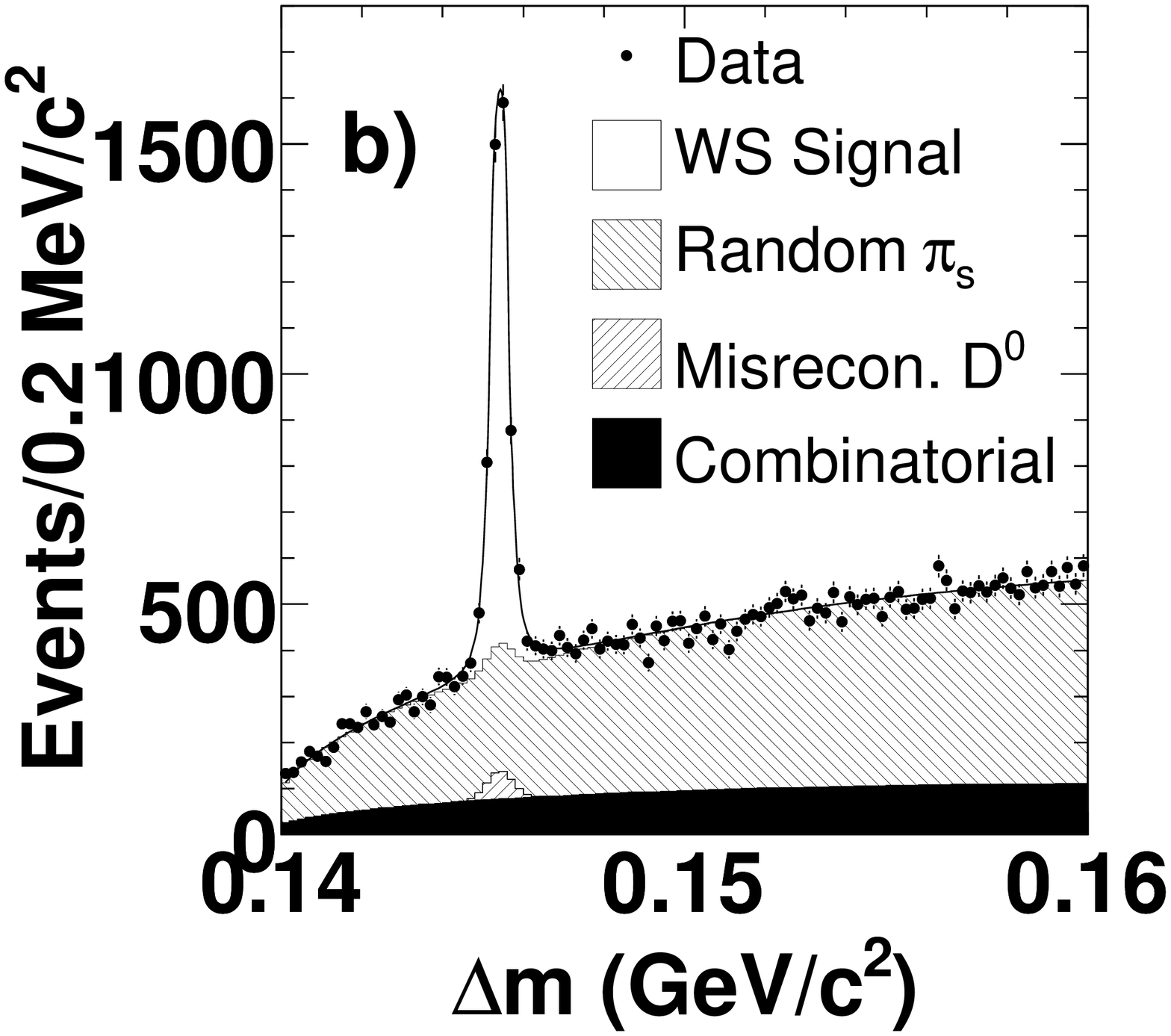}
  }
  \caption{a)  \mKpi for 
wrong-sign (WS) candidates with $0.1445<\dm<0.1465\gevcc$, and b)  \dm for WS candidates
with $1.843<\mKpi<1.883\gevcc$. The fitted PDFs are overlaid. 
The shaded regions represent the different background components.}
  \label{fig:r18Data_MdMfit}
  \smallskip
\end{figure}

We fit the RS and WS data samples simultaneously with shape parameters
describing the signal and random $\pisoft^+$ components shared between
the two data samples. We find $1,141,500\pm
1,200$ RS signal events and $4,030\pm 90$ WS signal events. The
dominant background component is the random $\pisoft^+$ background.
Projections of the WS data and fit are shown in Fig.~\ref{fig:r18Data_MdMfit}.

The measured proper-time distribution for the RS signal is described by an
exponential function convolved with a resolution function whose
parameters are determined by the fit to the data. The resolution function
is the sum of three Gaussians with widths
proportional to the estimated event-by-event proper-time uncertainty
\terr. The random $\pisoft^+$ background is described by the same proper-time
distribution as signal events, since the slow pion has little weight
in the vertex fit. The proper-time distribution of the combinatorial
background is described by a sum of two Gaussians, one of which has a
power-law tail to account for a small long-lived component.  The
combinatorial background and real \Dz decays have different
\terr distributions, as determined from data using a background-subtraction 
technique \cite{Pivk:2004ty} based on the fit to \mKpi and \dm.

The fit to the RS proper-time distribution is performed over all
events in the full \mKpi and \dm region. The PDFs for signal and
background in \mKpi and \dm are used in the proper-time fit with all
parameters fixed to their previously determined values.  The fitted
\Dz lifetime is found to be consistent with the world-average
lifetime~\cite{PDG2006}.

\begin{figure}[phtb]
  \centering
  \centerline{%
    \includegraphics[width=0.9\linewidth, clip=]{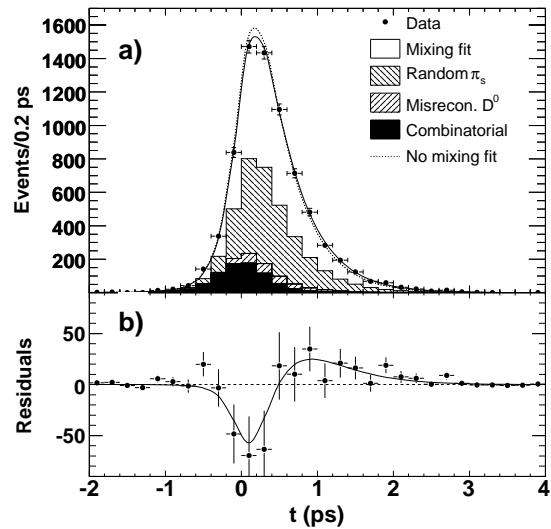}
  }
\caption{a) The proper-time distribution of combined \Dz and \Dzb 
WS candidates in the signal region
$1.843<\mKpi<1.883\gevcc$ and $0.1445<\dm<0.1465\gevcc$.
The result of the fit allowing (not allowing) mixing 
but not \CP violation 
is overlaid as a solid (dashed) curve.
Background components are shown as shaded regions.
b) The points represent the difference 
between the data and the no-mixing fit. The solid curve
shows the difference between fits with and without mixing.}
\label{fig:histTimeBiasWSR18Data}
\end{figure}

The measured proper-time distribution for the WS signal is modeled by
Eq.~(\ref{eq:Tws}) convolved with the resolution function determined in
the RS proper-time fit. The random $\pisoft^+$ and misreconstructed
\Dz backgrounds are described by the RS signal proper-time
distribution since they are real \Dz decays. 
The proper-time distribution for WS data is shown in
Fig.~\ref{fig:histTimeBiasWSR18Data}. The fit results with and without 
mixing are shown as the overlaid curves.

\begin{figure}[phtb]
  \centering
  \centerline{%
    \includegraphics[width=0.95\linewidth, clip=]{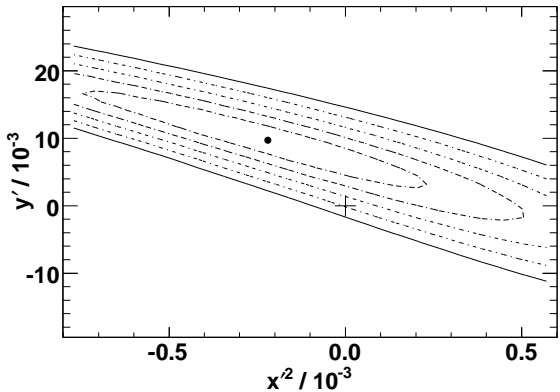}
}
\caption{The central value (point) and confidence-level (CL) contours for 
$1-\mbox{CL}=0.317\ (1\sigma)$, $4.55\times10^{-2}\ (2\sigma)$, 
$2.70\times10^{-3}\ (3\sigma)$, $6.33\times10^{-5}\ (4\sigma)$ and
$5.73\times10^{-7}\ (5\sigma)$, calculated from the change in the value
of $-2\ln{\cal L}$ compared with its value at the minimum.
Systematic uncertainties are included. The no-mixing point is shown 
as a plus sign~($+$).}
\label{fig:CPContour}
\end{figure}

The fit with mixing provides a substantially better description of the data
than the fit with no mixing.
The significance of the mixing signal is evaluated based on
the change in negative log likelihood with respect to the minimum.
Figure~\ref{fig:CPContour} shows confidence-level (CL) contours
calculated from the change in log likelihood ($-2\Delta\ln{\cal L}$) in two
dimensions (\xPrimeSq and \yPrime) with systematic uncertainties
included.  The likelihood maximum is at the unphysical value of
$\xPrimeSq=-2.2\times10^{-4}$ and $\yPrime = 9.7 \times 10^{-3}$. The
value of $-2\Delta\ln{\cal L}$ at the most likely point in the
physically allowed region ($\xPrimeSq=0$ and $\yPrime=6.4 \times
10^{-3}$) is $0.7$~units.  The value of
$-2\Delta\ln{\cal L}$ for no-mixing is $23.9$~units.
Including the systematic uncertainties, this corresponds to a
significance equivalent to 3.9~standard deviations
($1-\mbox{CL}=1\times10^{-4}$) and thus constitutes evidence for
mixing. The fitted values of the mixing parameters and \Rdcs are
listed in Table~\ref{tab:results}.  The correlation coefficient
between the \xPrimeSq and \yPrime parameters is $-0.94$.

Allowing for the possibility of \CP violation, we calculate the values
of $\Rdcs = \sqrt{\Rdcs^+\Rdcs^-}$ and 
$\AD = (\Rdcs^{+} - \Rdcs^{-})/(\Rdcs^{+} + \Rdcs^{-})$
listed in Table~\ref{tab:results}, from the fitted $\Rdcs^{\pm}$ values.
The best fit in each case is more than three
standard deviations away from the no-mixing hypothesis.
All cross checks indicate that the high level of agreement between
the separate \Dz and \Dzb fits is a coincidence.
\par
\begin{figure}[phtb]
  \centering
  \centerline{%
    \includegraphics[width=0.9\linewidth, clip=]{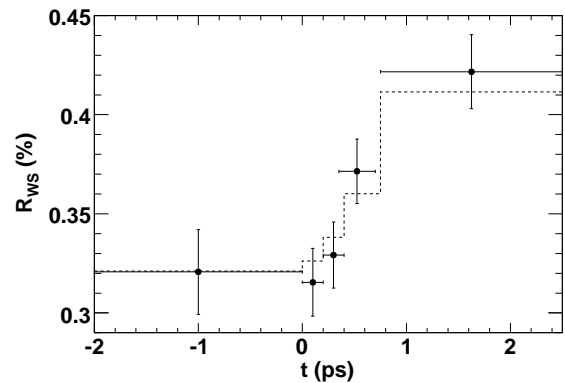}
  }
\caption{The WS branching fractions from independent \mdm\ fits to
slices in measured proper time (points). 
The dashed line shows the expected wrong-sign rate
as determined from the mixing fit shown in Fig.~\ref{fig:histTimeBiasWSR18Data}.
The $\chi^2$ with respect to expectation from the mixing fit is 1.5;
for the no-mixing hypothesis (a constant WS rate), the $\chi^2$ is 24.0.}
\label{fig:RwsTimeBins}
\end{figure}

As a cross-check of the mixing signal, we perform independent 
\mdm\ fits with no shared parameters for intervals in proper time selected
to have approximately equal numbers of RS candidates.
The fitted WS branching fractions are shown in
Fig.~\ref{fig:RwsTimeBins} and are seen to increase with time.  The
slope is consistent with the measured mixing parameters and
inconsistent with the no-mixing hypothesis.

\begin{table}[thb]
  \caption{Results from the different fits.
  The first uncertainty listed is statistical and the second systematic.}
  \label{tab:results}
  \centering\small
  \begin{ruledtabular}
    \begin{tabular}{lcr@{~$\pm$}r@{~$\pm$}r}
     Fit type & Parameter & \multicolumn{3}{c}{Fit Results ($/10^{-3}$)}  \\
    \hline
    No \CP viol. or mixing & $\Rdcs$ & $3.53 $ & $ 0.08 $ & $ 0.04$\\
    \hline
    \multirow{3}{1.7cm}{No \CP\\ violation}
    &  $\Rdcs$        & $3.03$ & $0.16$ & $ 0.10$ \\
    &  $\xPrimeSq$  & $-0.22$ & $0.30$ & $ 0.21$   \\
    &  $\yPrime$    & $9.7$ & $4.4$ & $ 3.1$      \\
    \hline
    \multirow{5}{1.7cm}{\CP\\ violation \\ allowed}
    & $\Rdcs$     & $3.03$ &$0.16$ & $0.10$  \\
    & $\AD$       & $-21$ & $52$ & $15$  \\
    & $\xPrimePSq$ & $-0.24 $ & $ 0.43 $ & $ 0.30 $\\
    & $\yPrimeP$   & $ 9.8  $ & $ 6.4  $ & $ 4.5  $\\
    & $\xPrimeMSq$ & $-0.20 $ & $ 0.41 $ & $ 0.29 $\\
    & $\yPrimeM$   & $ 9.6  $ & $ 6.1  $ & $ 4.3  $
  \end{tabular}
  \end{ruledtabular}
\end{table}
 
We have validated the fitting procedure on simulated data samples using
both MC samples with the full detector simulation and large
parameterized MC samples. In all cases we have found the fit to be
unbiased. As a further cross-check, we have performed a fit to the RS
data proper-time distribution allowing for mixing
in the signal component; the fitted values of the mixing parameters
are consistent with no mixing.  The correlations among parameters
determined at different stages of the fit are low.  In addition we have found the staged fitting approach to give the same solution and confidence regions as
a simultaneous fit in which all parameters are allowed to vary.


\par
In evaluating systematic uncertainties in \Rdcs and the mixing parameters
we have considered variations in the fit model and in the selection criteria.
We have also considered alternative forms of the \mKpi, \dm, proper time, and \terr
PDFs.  We varied the $t$ and \terr requirements.
In addition, we considered 
variations that keep or reject all
\Dstp\ candidates sharing tracks with other candidates. 
\par
For each source of systematic error, we compute the significance
$\signif_i^2=2\left[\ln\like\xPSQyP-\ln\like(\xPrimeSq_i, \yPrime_i)\right]/2.3$, where 
\xPSQyP are the parameters obtained from the standard fit, 
$(\xPrimeSq_i, \yPrime_i)$ the parameters from the fit including the $i^{th}$ systematic variation,
and \like\ the likelihood of the standard fit. 
The factor 2.3 is the 68\% confidence level for 2 degrees of freedom. 
To estimate the significance of our results in \xPSQyP, 
we reduce $-2\Delta\ln{\cal L}$ by a factor of $1+\Sigma s_i^2=1.3$
to account for systematic errors. The largest contribution to this factor, $0.06$, is due to uncertainty in modeling
the long decay time component from other $D$ decays in the signal region. The second largest 
component, $0.05$, is due to the presence of a non-zero mean in the proper time signal resolution
PDF.
The mean value is determined in the RS proper time fit to be $3.6\fs$ and is due to
small misalignments in the detector.
The
error of $15\times 10^{-3}$ on \AD is primarily due to uncertainties in modeling the differences
between \Kp and \Km absorption in the detector.

\par
We have presented evidence for $\Dz$-$\Dzb$ mixing.
Our result is inconsistent with the no-mixing hypothesis 
at a significance of 3.9 standard deviations. 
We measure $\yPrime=[9.7 \pm 4.4 \hbox{ (stat.)}\pm 3.1 \hbox{ (syst.)}] \times 10^{-3}$,
while \xPrimeSq is consistent with zero. We find no evidence for \CP
violation and measure \Rdcs to be $[0.303\pm0.016\hbox{ (stat.)}\pm 
0.010\hbox{ (syst.)}]\%$.  The result is consistent with
SM estimates for mixing.

\begin{acknowledgments}
We are grateful for the excellent luminosity and machine conditions
provided by our \pep2\ colleagues, 
and for the substantial dedicated effort from
the computing organizations that support \babar.
The collaborating institutions wish to thank 
SLAC for its support and kind hospitality. 
This work is supported by
DOE
and NSF (USA),
NSERC (Canada),
IHEP (China),
CEA and
CNRS-IN2P3
(France),
BMBF and DFG
(Germany),
INFN (Italy),
FOM (The Netherlands),
NFR (Norway),
MIST (Russia),
MEC (Spain), and
PPARC (United Kingdom). 
Individuals have received support from the
Marie Curie EIF (European Union) and
the A.~P.~Sloan Foundation.

\end{acknowledgments}

\bibliography{babar-pub-0719}

\begin{thebibliography}{22}
\bibitem[{foo()}]{footnote:CC}
\bibinfo{note}{The use of charge-conjugate modes is implied unless otherwise
  noted.}
\end{thebibliography}

\end{document}